\documentstyle[11pt,newpasp,twoside,epsf]{article}
\input psfig.tex

\def\edcomment#1{\iffalse\marginpar{\raggedright\sl#1\/}\else\relax\fi}
\reversemarginpar

\def\al{$\alpha$}

\def\apj{{ApJ}}
\def\apjs{{ApJS}}

\def\lamb{$\lambda$}
\def\lax{{$\mathrel{\hbox{\rlap{\hbox{\lower4pt\hbox{$\sim$}}}\hbox{$<$}}}$}}
\def\gax{{$\mathrel{\hbox{\rlap{\hbox{\lower4pt\hbox{$\sim$}}}\hbox{$>$}}}$}}
\def\simlt{\lower.5ex\hbox{$\; \buildrel < \over \sim \;$}}
\def\simgt{\lower.5ex\hbox{$\; \buildrel > \over \sim \;$}}
\def\lum{erg s$^{-1}$}

\def\percm2{cm$^{-2}$}

\def\solmass{$M_\odot$}

\begin{document}
\title{The Central Engines of Low-Luminosity AGNs}
 \author{Luis C. Ho}
\affil{The Observatories of the Carnegie Institution of Washington, 813 Santa 
Barbara Street, Pasadena, CA 91101, U.S.A.}

\begin{abstract}
I summarize the main characteristics of AGNs in nearby galaxies 
and present a physical picture of their central engines.
\end{abstract}

\vspace*{-0.5cm}
\section{The ``Top 10'' Properties of Low-luminosity AGNs}

Although the physical nature of some low-luminosity AGNs (LLAGNs) is still not 
fully understood, the weight of the recent cumulative evidence suggests that a 
significant fraction of them are genuinely accretion-powered sources.  Here 
I identify the most important observational characteristics of these objects, 
which point to some novel insights on the structure of their central engines.

\vspace*{0.1cm}
\noindent
{\it (1) Demography.}\ LLAGNs are very common.  According to the Palomar 
survey (Ho, Filippenko, \& Sargent 1997), over 40\% of nearby galaxies, and an 
even greater fraction of bulge-dominated systems, contain LLAGNs.

\vspace*{0.1cm}
\noindent
{\it (2) Low ionization.}\ The dominant population (2/3) of LLAGNs have 
low-ionization state spectra.  They are classified as either LINERs or 
transition objects.

\vspace*{0.1cm}
\noindent
{\it (3) Low accretion power.}\  LLAGNs are intrinsically faint.  
Figure~1{\it a}, from Ho (2003), shows the distributions of bolometric 
luminosities for $\sim$250 objects from the Palomar survey.  Note that nearly 
all the objects have $L_{\rm bol} < 10^{44}$ \lum, and most significantly 
less. Seyferts are on average 10 times more luminous than LINERs or transition 
objects.

\vspace*{0.1cm}
\noindent
{\it (4) Sub-Eddington.}\  LLAGNs are highly sub-Eddington systems, as shown 
in Figure~1{\it b}.  Most have $\lambda \equiv L_{\rm bol}/L_{\rm Edd} < 
10^{-2}$.  Seyferts have systematically lower $\lambda$ than LINERs or 
transition objects.

\vspace*{0.1cm}
\noindent
{\it (5) Radiatively inefficient.}\  Direct measurements of accretion rates are
not available, but rough estimates can be made of the likely minimum 
rates supplied {\it in situ}\ through stellar mass loss and Bondi capture of 
hot gas.  Ho (2003) finds $\dot M$ \gax\ $10^{-5} - 10^{-3}$ \solmass\ 
yr$^{-1}$.  If this gas were to be all accreted and radiates with a standard 
efficiency of $\eta = 10$\%, the nuclei should be $1-4$ orders of magnitude 
more luminous than observed.  This suggests that either only a tiny fraction 
of the available gas gets accreted (the rest, e.g., being driven out by winds) 
or that $\eta$ is much less than 10\%.  The latter possibility is consistent 
with models for radiatively inefficient accretion flows (e.g., Quataert 2001). 
(Incidentally, the above estimates indicate that, insofar as the majority of 
nearby AGNs are concerned, there is no motivation to seek additional sources 
of fuel supply, such as through bar dissipation.  Far from needing to find 
ways to feed the nucleus from gas on large scales, the problem is in fact the 
opposite, namely how to get rid of or hide the material that is already 
there.)

\clearpage

%Figure 1
%80 70 400 940
\begin{figure}
\vbox{
\hbox{
\hskip 0truein
\psfig{file=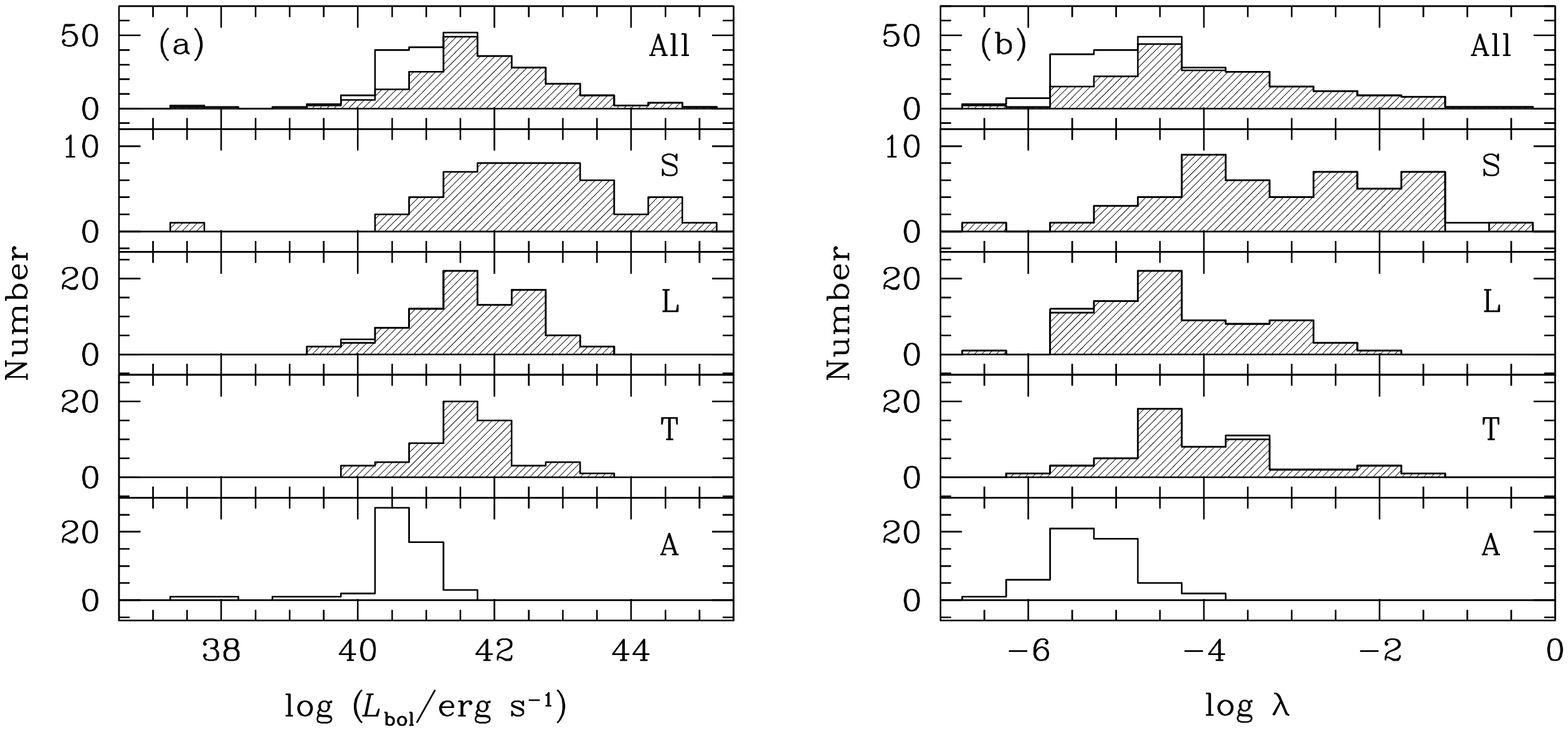,height=2.5truein,angle=270}
}
}
\vskip +0.7cm
\noindent{Fig. 1.} Distribution of ({\it a}) nuclear bolometric luminosities 
and ({\it b}) Eddington ratios $\lambda\,\equiv\,L_{\rm bol}/L_{\rm Edd}$.
S = Seyferts, L = LINERs, T = transition objects, and A = absorption-line 
nuclei.  Open histograms denote upper limits.  From Ho (2003.)
\end{figure}

\vspace*{-0.9cm}

\vspace*{0.1cm}
\noindent
{\it (6) No ``big blue bump.''}\   With few exceptions, the spectral energy 
distributions (SEDs) of LLAGNs lack the optical--UV ``big blue bump,'' a 
feature usually attributed to thermal emission from an optically thick, 
geometrically thin accretion disk (Ho 1999; Ho et al. 2000).  Instead, 
there appears to be an IR excess.

\vspace*{0.1cm}
\noindent
{\it (7) ``Big red bump.''}\   Most of the SEDs contain a maximum in the 
IR, whose peak is currently poorly defined because of the current lack of 
sufficient high-resolution IR data.

\vspace*{0.1cm}
\noindent
{\it (8) Radio loud.}\   The SEDs of LLAGNs are also generically radio loud.
This is true of most LINERs (Ho 1999, 2002; Ho et al. 2000; Terashima \& 
Wilson 2003), and, contrary to persistent popular misconception,  is true even 
for most Seyfert nuclei (Ho \& Peng 2001).  

\vspace*{0.1cm}
\noindent
{\it (9) No broad Fe K\al\ line.}\  The 6.4 keV Fe K\al\ line is detected in 
some LLAGNs, but it is almost always narrow (Terashima et al. 2002).

\vspace*{0.1cm}
\noindent
{\it (10) ``Batman'' line profiles.}\  Emission lines with broad, 
double-peaked profiles, taken to be the kinematic signature of a 
relativistically broadened disk, are found quite often in LLAGNs (Ho et al. 
2000, and references therein; Barth et al.  2001; Eracleous \& Halpern 2001).  

%Figure 2
%1 1 560 263
\begin{figure}
\vbox{
\hbox{
\hskip 0.2truein
\psfig{file=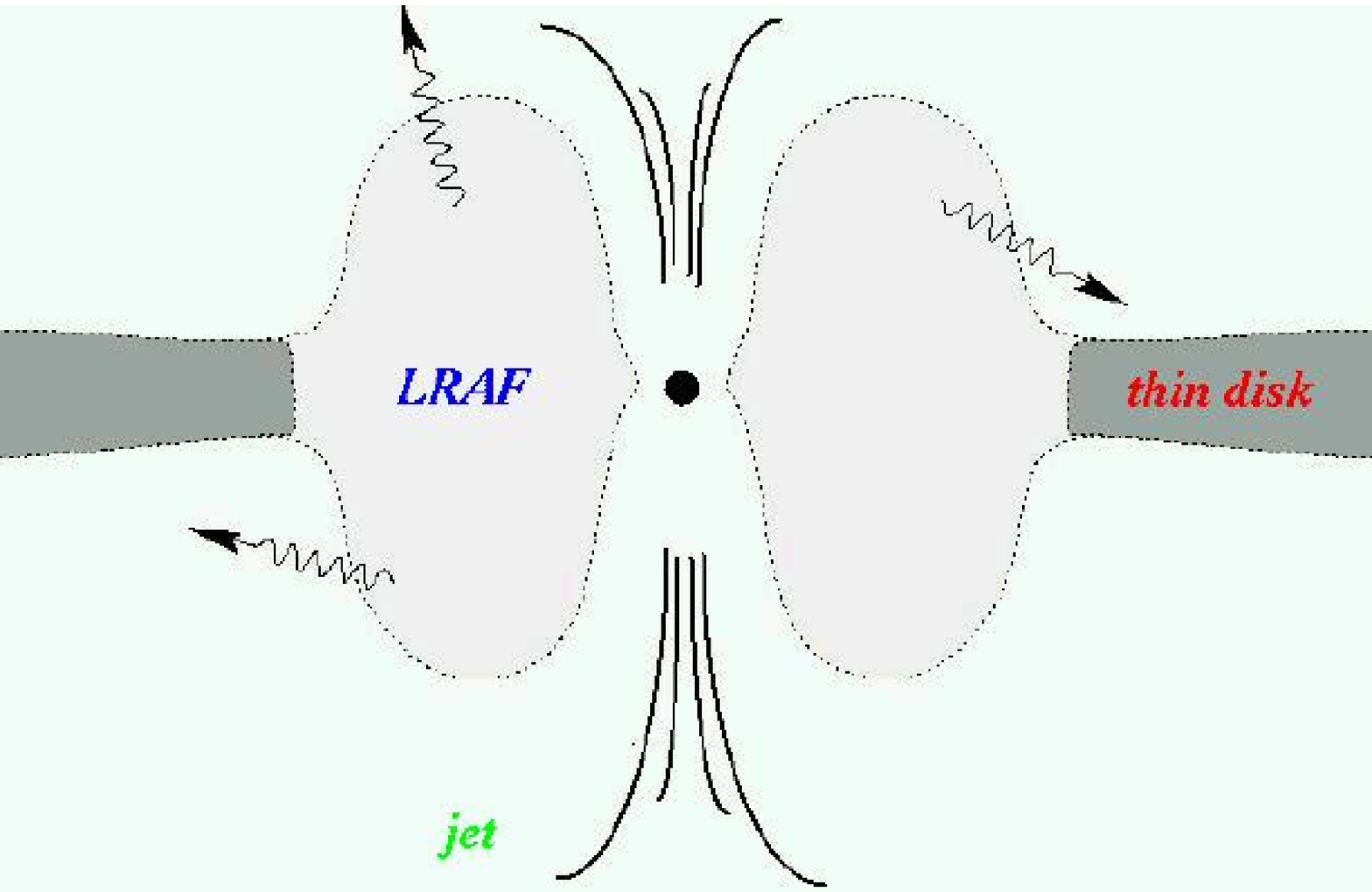,height=3.2truein,angle=0}
}
}
\vskip +0.5cm
\noindent{Fig. 2.} A cartoon depicting the structure of the accretion flow
surrounding weakly active massive black holes.  An inner low-radiative 
efficiency accretion flow (LRAF) irradiates an outer thin disk.  Jets may be 
present.
\end{figure}

\vspace*{-0.3cm}
\section{A Physical Picture of the Central Engine}
\vspace*{-0.2cm}

I propose that the above set of characteristics, common to most LLAGNs 
studied in detail thus far, suggest that nearby galaxy bulges contain 
central engines as schematically depicted in Figure~2.  Most galaxies 
with bulges contain active nuclei because most, if not all, bulges contain 
massive black holes.  This is the picture that is emerging from recent 
kinematical studies of nearby galaxies (e.g., Gebhardt et al. 2003).  In the 
present-day universe, and especially in the centers of big bulges, the amount 
of gas available for accretion is quite small, plausibly well below the 
Eddington rate for the associated black hole mass (Ho 2003).  In such a 
regime, the low-density, 
tenuous material is optically thin and cannot cool efficiently.  Rather 
than settling into a classical optically thick, geometrically thin disk, 
the hot accretion flow assumes a quasi-spherical configuration, whose dynamics 
may be dominated by advection, convection, or outflows (see Quataert 2001 for 
a review).  For simplicity, I follow Quataert (2001) and simply call these 
low-radiative efficiency accretion flows (LRAFs).  The existence of LRAFs 
in these systems, or conversely the absence of classical thin disks extending 
all the way to small radii (few $R_S$), is suggested by their (1) low 
luminosities, (2) low Eddington ratios, (3) low inferred radiative 
efficiencies, (4) lack of a big blue bump, and (5) lack of relativistically 
broadened Fe K\al\ lines.  

Apart from a central LRAF, two additional components generally seem to be 
required.  First, detailed considerations of the broadband SED show that the 
baseline LRAF spectrum underpredicts the observed radio power (e.g., Quataert 
et al. 1999; Ulvestad \& Ho 2001).  Most of the radio luminosity, which is 
substantial because these objects tend to be ``radio-loud,''  must come 
from another component, and the most likely candidate is a compact jet.   Does 
the puffed-up structure of an LRAF, or its propensity for outflows, somehow 
facilitate the generation of relativistic jets?  Second, an {\it outer}\ 
thin disk, truncated at perhaps $\sim 100-1000\, R_S$,  seems necessary to 
explain (1) the existence of the ``big red bump'' in the SED (e.g., Quataert 
et al. 1999) and (2) the prevalence of double-peaked broad emission lines 
(Chen, Halpern, \& Filippenko 1989; Ho et al. 2000).

Lastly, we note that low-ionization spectra may emerge quite naturally in the 
scenario suggested above.  In the context of AGN photoionization models, it is 
well known that LINER-like spectra can be produced largely by lowering the 
``ionization parameter'' $U$, typically by a factor of $\sim 10$ below that in 
Seyferts (e.g., Halpern \& Steiner 1983; Ferland \& Netzer 1983).  The 
characteristically low luminosities of LINERs (Fig.~1{\it a}), coupled 
with their low densities (Ho, Filippenko, \& Sargent 2003), naturally lead 
to low values of $U$.  Two other effects, however, are also important in 
boosting the low-ionization lines.  All else being equal, hardening the 
ionizing spectrum (by removing the big blue bump) in photoionization 
calculations creates a deeper partially ionized zone from which low-ionization 
transitions, especially [O~I] \lamb\lamb 6300, 6363, are created.  
Because of the prominence of the radio spectrum, cosmic-ray heating of the 
line-emitting gas by the radio-emitting plasma may be nonnegligible; one 
consequence of this process is again to enhance the low-ionization lines 
(Ferland \& Mushotzky 1984).

\vspace*{0.1cm}

\acknowledgments{
L. C. H. acknowledges financial support from the Carnegie Institution of 
Washington and from NASA grants.

\end{document}